# A degenerate trion liquid in atomic double layers

Phuong X. Nguyen[1], Raghav Chaturvedi[1], Liguo Ma[1], Patrick Knuppel[1], Kenji Watanabe[2], Takashi Taniguchi[2], Kin Fai Mak[1,3,4*], Jie Shan[1,3,4*]

[1]School of Applied and Engineering Physics, Cornell University, Ithaca, NY, USA
[2]National Institute for Materials Science, Tsukuba, Japan
[3]Laboratory of Atomic and Solid State Physics, Cornell University, Ithaca, NY, USA
[4]Kavli Institute at Cornell for Nanoscale Science, Ithaca, NY, USA

[*]Email: kinfai.mak@cornell.edu; jie.shan@cornell.edu

These authors contributed equally: Phuong X. Nguyen, Raghav Chaturvedi

**Abstract**
Trions are a three-particle bound state of electrons and holes. Experimental realization of a trion liquid in the degenerate quantum limit would open a wide range of phenomena in quantum many-body physics. However, trions have been observed only as optically excited states in doped semiconductors to date. Here we report the emergence of a degenerate trion liquid in a Bose-Fermi mixture of holes and excitons in Coulomb-coupled $MoSe_2$/$WSe_2$ monolayers. By electrically tuning the hole density in $WSe_2$ to be two times the electron density in $MoSe_2$, we generate equilibrium interlayer trions with binding energy about 1 meV at temperatures two orders of magnitude below the Fermi temperature. We further demonstrate a density-tuned phase transition to an electron-hole plasma, spin-singlet correlations for the constituent holes and Zeeman-field-induced dissociation of trions. The results pave the way for exploration of the correlated phases of composite particles in solids.

**Main**

Quantum liquids of coexisting oppositely charged particles are an ideal platform to generate quasiparticles with distinct quantum statistics and interactions and to realize proposed many-body phases of matter. Examples include exciton condensates (*1-18*), exciton-mediated superconductivity (*19-22*), Wigner crystals of excitons and trions (*23, 24*) and supersolids (*25, 26*). While it is straightforward to generate quantum fluids of electrons (e) or holes (h) in a solid-state system, their mixtures are intrinsically short-lived due to e-h recombination (*27, 28*). The recent development of strongly Coulomb-coupled double layers of transition metal dichalcogenide (TMD) semiconductors (*14, 15, 17, 18, 29, 30*) has overcome this experimental challenge. In these devices, metal electrodes can separately contact the two semiconductor layers, and inject electrons into one layer and holes into the other layer with independently tunable densities (*6, 7, 13, 31*). This advance has enabled both the thermodynamic (*15, 30*) and transport studies (*17, 18*) of strongly correlated excitons (a composite boson consisting of a bound e-h pair) in thermal equilibrium. This advance has also opened the door to realize a Bose-Fermi mixture and new quasiparticles by introducing additional electrons or holes to the excitons.

Here, we report the emergence of a degenerate trion liquid in Coulomb-coupled MoSe$_2$/WSe$_2$ double layers when the hole density ($p$) is two times the electron density ($n$) (Fig. 1A, B). By combining transport, capacitance and optical measurements, we explore the properties of equilibrium trions (binding energy of ~ 1 meV and spin-singlet correlation for two constituent holes). We also demonstrate a density-tuned quantum phase transition from a degenerate trion liquid to an e-h plasma and dissociation of trions by a Zeeman field of several Teslas. Our results pave the way for exploration of quantum many-body phenomena of exciton-mediated pairing (*19-22*), unconventional trion and Fermi polaron metals (*32*), and trion-to-polaron crossover (*33, 34*).

We employ the dual-gated device structure shown in Fig. 1C. The details have been reported previously in Ref. (*17*). The MoSe$_2$ (Mo) and WSe$_2$ (W) monolayers are separated by a thin hexagonal boron nitride (hBN) dielectric layer (1.5 - 2 nm in the channel). An interlayer bias voltage ($V_b$) tunes the electron-hole pair density ($n_{pair}$) with electrons residing in the Mo layer and holes in the W layer (Fig. 1A). The relatively thick hBN spacer, together with the intentionally angle-misaligned Mo and W layers, has been shown to effectively suppress interlayer tunneling so that the excitons and trions are in thermal equilibrium (see Fig. S8). At the same time, the spacer is sufficiently thin so that the electrons and holes are in the strong coupling regime. To complete the device, the e-h double layer is embedded in nearly symmetric top and bottom gates made of hBN dielectrics and few-layer graphite electrodes. The symmetric combination of the gate voltages (denoted as $V_g$) controls the net charge density (or e-h imbalance); and the antisymmetric combination (denoted as $\Delta$) tunes the perpendicular electric field, which can heavily dope the semiconductors in the contact region and turn on the contacts.

We perform standard four-terminal resistance measurements on the W layer (because of its low contact resistances) while keeping the Mo layer in the open-circuit configuration (Fig. 1D). No current is allowed to flow in the Mo layer. When the Mo layer is charge neutral ($n = 0$), the four-terminal resistance $R$ probes the electrical transport properties of monolayer WSe$_2$. But when the Mo layer is electron-doped, $R$ becomes sensitive to interlayer Coulomb interactions. Specifically, $R$ is expected to exhibit characteristics of an insulating state with a charge gap at low temperatures when interlayer bound states such as trions form. The charge gap characterizes the binding energy. We also measure the penetration capacitance (*15*) to verify that the trion fluid (that is, the e-h double layer as a whole) is charge compressible. To characterize the spin configuration of the trion ground state, we perform magnetic circular dichroism (MCD) measurements. A spin-singlet correlation of the constituent holes in the trion is expected to give rise to suppressed magnetic susceptibility of the W layer. Hence an external Zeeman field can dissociate trions by overcoming the spin gap and induce an insulator-to-metal transition in the W layer. We have studied 3 devices (Fig. S1 and S3) and all results are reproduced. See Supplementary Materials for details on device fabrication and electrical and MCD measurements.

**Degenerate liquid of trions**

Figure 1E and 1F show the penetration capacitance ($C_P/C_{gg}$, normalized by the gate-to gate geometrical capacitance, $C_{gg}$) and the four-terminal resistance $R$ of the W layer as a function of $V_b$ and $V_g$ at 1.5 K ($\Delta$ is fixed at 5.5 V). The penetration capacitance is proportional to the electronic incompressibility of the double layer (*15*). Specifically, $C_P/C_{gg} \approx 1$ ($\approx 0$) corresponds to an incompressible (compressible) state. Similar to earlier reports (*15, 17, 18, 30*), we can identify four distinct regions in the electrostatics phase diagram for the W and Mo layers: *pn*, *pi*, *ii* and *in*, where *p*, *n* and *i* denote the hole-doped, electron-doped and charge neutral regions, respectively. The boundary between the *pi*- and *ii*-regions does not disperse with $V_b$ because the W layer is grounded. In addition, a charge-incompressible region of triangular shape in the *pn*-region is observed; it corresponds to the excitonic insulator phase with $n = p$.

The four-terminal resistance $R$ of the W layer is largely consistent with the electrostatics phase diagram. We observe small resistances when the W layer is doped (the *pn*- and *pi*-regions) with two exceptions. The first is the excitonic insulator phase. The observed large resistance here is consistent with the recent report of perfect Coulomb drag and diverging drag resistance at low temperatures from the formation of interlayer bound e-h pairs (*17, 18*). The second exception is the state that stems from the left lower corner of the excitonic insulator phase and follows the $p = 2n$ line (see Supplementary Materials and Fig. S2 for density calibration). This indicates the emergence of a new interlayer bound state.

We examine the temperature dependence of the state around $p = 2n$ in Fig. 2. At lattice temperature 0.02 K, the state becomes more prominent in the four-terminal resistance map in Fig. 2A. We take a linecut at a fixed $V_g$ (red dashed), which corresponds to approximately constant net positive charge density $2.42 \times 10^{11}$ cm$^{-2}$, and show resistance as a function of $V_b$ (approximately e-h pair density) at temperature ranging from 0.02 K to 10 K in Fig. 2B. When the Mo layer is charge neutral ($n = 0$) at small $V_b$, we observe a metallic behavior, that is, $R$ increases with increasing temperature. This behavior is expected for hole-doped monolayer WSe$_2$ beyond the mobility edge (at $p \approx 1.8 \times 10^{11}$

cm$^{-2}$, Fig. S7). When $V_b$ exceeds a threshold at about 0.86 V, e-h pairs are injected into the double layer; $R$ first decreases due to the increased hole density then increases dramatically. Around $p = 2n$ (shaded region), $R$ displays an insulating behavior, that is, $R$ increases with decreasing temperature. Upon further increases of the e-h pair density ($n > p/2$), the resistance again exhibits a metallic behavior.

The insulating behavior of the W layer upon an increase of the hole density (through injection of e-h pairs) in the shaded region cannot be explained by either single-particle physics or disorder. Disorder effects are inconsistent with the resistance drop observed at lower hole densities (immediately left to the shaded region) since disorder potentials are expected to be more effectively screened at higher hole densities (*35*). The insulating state emerges only around $p = 2n$. This shows that all the holes in the W layer are bound to the electrons in the Mo layer in a 2:1 ratio at low temperatures, providing strong evidence for the formation of positive trions.

We characterize the properties of trions by tracing the four-terminal resistance along the $p = 2n$ line. We show the trion density dependence of $R$ at varying temperatures in Fig. 2C and the temperature dependence of $R$ at varying trion densities in Fig. 2D. A metal-insulator transition is observed at critical trion density $3.75 \times 10^{11}$ cm$^{-2}$, where $R$ is nearly temperature independent (dashed line in Fig. 2D). The system is insulating below and metallic above the critical density. We estimate the charge gap of the insulating state by performing a thermal activation fit to the temperature dependence of $R$ (Fig. S4). The density dependence of the gap size is shown in the inset of Fig. 2C. The charge gap decreases monotonically with increasing trion density and vanishes continuously at the critical density.

The charge gap is a measure of the trion binding energy. It is about 1 meV at trion density of ~$2 \times 10^{11}$ cm$^{-2}$, in comparison to 20-30 meV for intralayer trions in TMD monolayers (*28*). The corresponding Fermi temperature is estimated to be ~ 5 K (Supplementary Materials). The trion liquid (strongly interacting) is thus in the degenerate limit in our experiment with temperature up to about two orders of magnitude below the Fermi temperature. The trion liquid is also charge compressible as shown by the penetration capacitance result (Fig. 1E); any injected holes are readily accepted by the exciton reservoir to form trions. Our result (inset, Fig. 2C) further suggests a density-tuned continuous phase transition from a trion liquid to an e-h plasma at the critical trion density. In the latter, excitons are unlikely to exist because the trion critical density is comparable to the exciton Mott density of $\approx 4 \times 10^{11}$ cm$^{-2}$ [Ref. (*17*)].

We note that the negative trion state is also expected to emerge from the right lower corner of the excitonic insulator phase around $n = 2p$. It is not observed in the current transport data because of the large resistance background. The W layer is highly resistive in this region due to the overall low hole densities and strong excitonic binding of the holes to the electrons in the Mo layer. (The negative trion state can be best accessed by four-terminal resistance measurements on the Mo layer.) Signatures of the negative trion state are observed in the complementary optical measurement (Fig. S6).

## Spin-singlet correlation in trions

Next we examine the spin configuration of trions by performing magneto-optical measurements (Supplementary Materials and Fig. S6). Figure 3A shows the magnetic circular dichroism (MCD) of the W layer as a function of $V_b$ and $V_g$ under out-of-plane magnetic field $B = 0.2$ T at 1.6 K. The MCD (spectrally integrated MCD over the intralayer charged exciton resonance) has been shown to be proportional to the spin (locked with valley) polarization of doped carriers in monolayer TMDs (*36*). In the small field limit, MCD is proportional to the magnetic susceptibility (*36*). We observe significantly enhanced magnetic susceptibility of the W layer in the excitonic insulator phase, suggesting the possibility of a spontaneously valley-polarized ground state at low temperatures as predicted by mean-field calculations (*7*). We also observe a susceptibility enhancement for small hole densities (boundary between the *pi*- and *ii*-regions), where the hole band mass and g-factor are renormalized (*37, 38*). The most intriguing observation is the susceptibility suppression around $p = 2n$, which fades away with increasing trion density above the critical value determined above in the transport study.

Figure 3C displays the temperature dependence of susceptibility for three representative values of $V_b$ at a fixed $V_g$ (red dashed line in Fig. 3A), corresponding to $n = 0$, $= p/2$ and $> p/2$, respectively. At high temperatures ($T \gtrsim 10$ K), the susceptibility is comparable in all three cases. At low temperatures, the susceptibility for the case of $p = 2n$ is strongly suppressed compared to the other two cases, for which the susceptibility rapidly increases with decreasing temperature. Figure 3B shows MCD along the red dashed in Fig. 3A extended up to 6 T. The MCD around $p = 2n$ remains small up to about 3-4 T, above which it rapidly saturates.

The substantially suppressed low-temperature susceptibility around $p = 2n$ demonstrates a spin-singlet correlation for the two holes in a positive trion. This is consistent with earlier optical studies of monolayer TMDs, where the ground state trions are a spin-singlet (*28, 39*) (i.e. two holes occupy different valley/spin states). The spin-singlet correlation is important only below the trion ionization temperature when a substantial population of trions is stabilized. This temperature scale (10 K) is consistent with the trion binding energy (1 meV) at $n_{trion} \approx 1.9 \times 10^{11}$ cm$^{-2}$. The spin-singlet correlation can also be overcome by an external Zeeman field when the Zeeman energy is comparable to the trion binding energy. The observed Zeeman field scale (3-4 T) is also consistent with the trion binding energy (Fig. 2C inset), using the reported hole g-factor $g \approx 10$ [Ref. (*37*)].

## Zeeman-field-induced dissociation of trions

Finally, we examine the fate of the trion liquid after the Zeeman field overcomes the spin-singlet correlation. Figure 4A shows the four-terminal resistance $R$ as a function of $B$ and $V_b$ at a fixed $V_g$, which corresponds to approximately constant net positive charge density $2.4 \times 10^{11}$ cm$^{-2}$. We observe three distinct sets of quantum oscillations. The first (horizontal lines below the e-h pair injection threshold, $n = 0$) is the quantum oscillations of monolayer WSe$_2$. They are observable above about 2 T, which corresponds to a quantum mobility exceeding 5,000 cm$^2$V$^{-1}$s$^{-1}$. The second set (nearly horizontal lines at large $V_b$) may arise from trions and its origin requires further studies. The main feature of interest is the Landau fan that emerges from $V_b = 0.866$ V (corresponding to $p = 2n$) above critical field ~ 5 T. The Landau fan disperses with both $V_b$ and $B$.

Figure 4B displays the magnetic-field dependence of $R$ for $p = 2n$ at varying temperatures. We observe a magnetic-field-induced metal-insulator transition at the critical field. The system is insulating below and metallic above the critical field. The critical field also corresponds to the magnetic saturation in Fig. 3B. Figure 4C shows the same study for a higher trion density. It shows the same properties but with a lower critical magnetic field.

The metal-insulator transition shows that once the Zeeman field overcomes the spin-singlet correlation and polarizes all holes to a single valley in the W layer, the trion liquid is dissociated into an e-h plasma. This is consistent with the observed smaller critical field for higher trion density since the trion binding energy decreases with increasing density (Fig. 2C inset), and a smaller trion binding energy requires a smaller Zeeman field to overcome the spin-singlet correlation. The result is inconsistent with partial ionization to yield a mixture of holes and dipolar excitons because in this scenario, the hole Landau levels are not expected to disperse with $V_b$ (which tunes the e-h pair density). The complete ionization of trions is presumably due to the substantial state-filling and free-carrier screening effects (*28, 39*) after magnetic saturation. Further theoretical studies are required to fully understand the field-induced trion dissociation process.

**Acknowledgements**
We thank Allan H. MacDonald, Yongxin Zeng and Bo Zou for helpful discussions.

# Figures

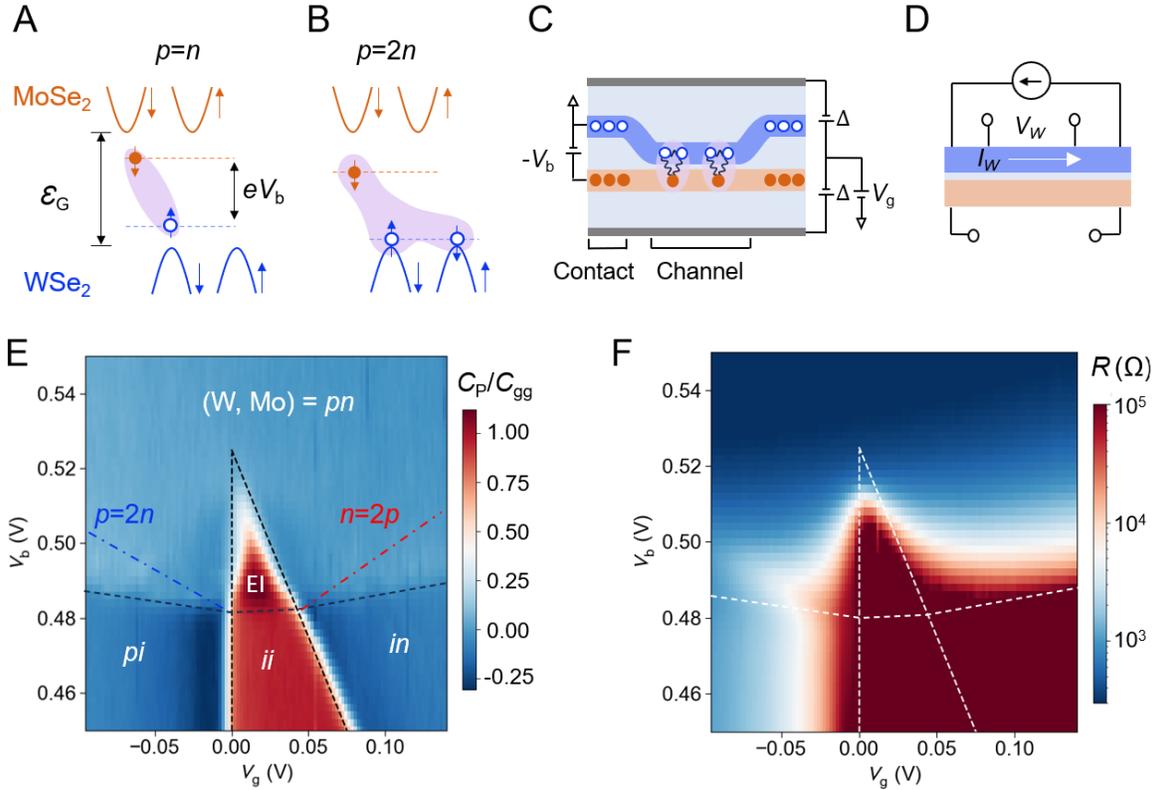

**Figure 1. Coulomb-coupled electron-hole double layers. A,B,** Schematic illustration of excitons (**A**) and trions (**B**) for one of the species. The conduction (valence) band minimum (maximum) resides in the MoSe$_2$ (WSe$_2$) layer. Each band is spin-valley degenerate with electron spins shown by arrows. The trion ground state is a spin-singlet for two holes. Dashed lines are the chemical potential of electrons and holes. $\mathcal{E}_G$ is the double layer band gap and $eV_b$ is the exciton chemical potential. **C,** Device schematic. The electron (orange) and hole (blue) layers are separated by 1.5-2 nm in the channel and 10-20 nm in the contact region. $V_b$, $V_g$ and $\Delta$ are the interlayer bias voltage, the symmetric and antisymmetric parts of the gate voltages, respectively. **D,** Four-terminal resistance $R = I_W/V_W$ is measured in the hole layer with the electron layer in open circuit, where $I_W$ and $V_W$ are the bias current and voltage drop, respectively. **E,F,** Normalized penetration capacitance $C_P/C_{gg}$ (**E**) and four-terminal resistance $R$ (**F**) versus $V_g$ and $V_b$ at 1.5 K (device 1). $C_{gg}$ is the gate-to-gate geometrical capacitance. Dashed lines denote the phase boundaries of hole-doped ($p$), electron-doped ($n$) and intrinsic ($i$) W and Mo layers. The electron-hole density ratio is 1:2 and 2:1 along the dash-dotted lines. EI denotes the excitonic insulator region inside the $pn$ regime.

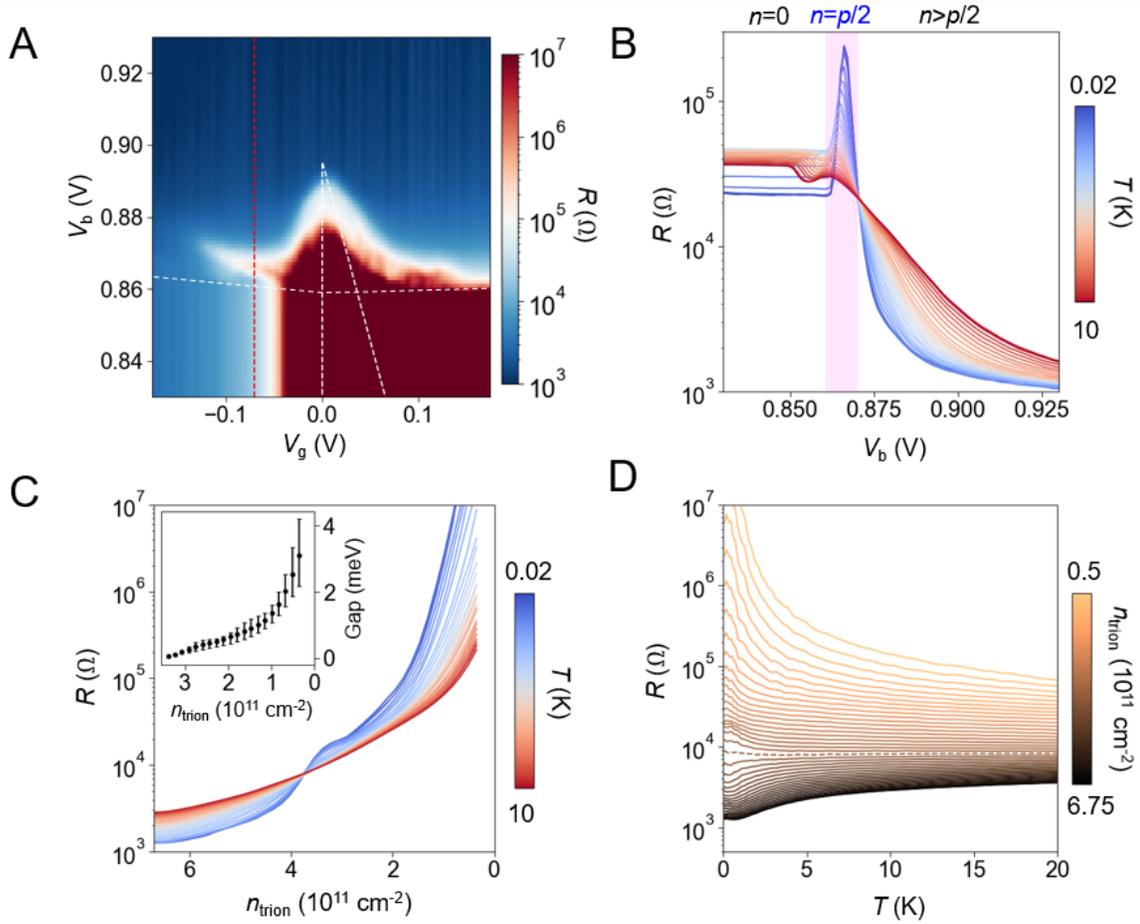

**Figure 2. Degenerate trion liquid. A,** Four-terminal resistance $R$ versus $V_b$ and $V_g$ at lattice temperature 20 mK (device 2). White dashed lines denote the electrostatics phase boundaries. **B,** Linecut of **A** along the red dashed line at varying temperatures (0.02-10 K). **C,D,** Four-terminal resistance $R$ along the $p = 2n$ line as a function of trion density at varying temperatures (**C**) and as a function of temperature at varying trion densities (**D**). The inset in **C** shows the estimated transport gap as a function of trion density.

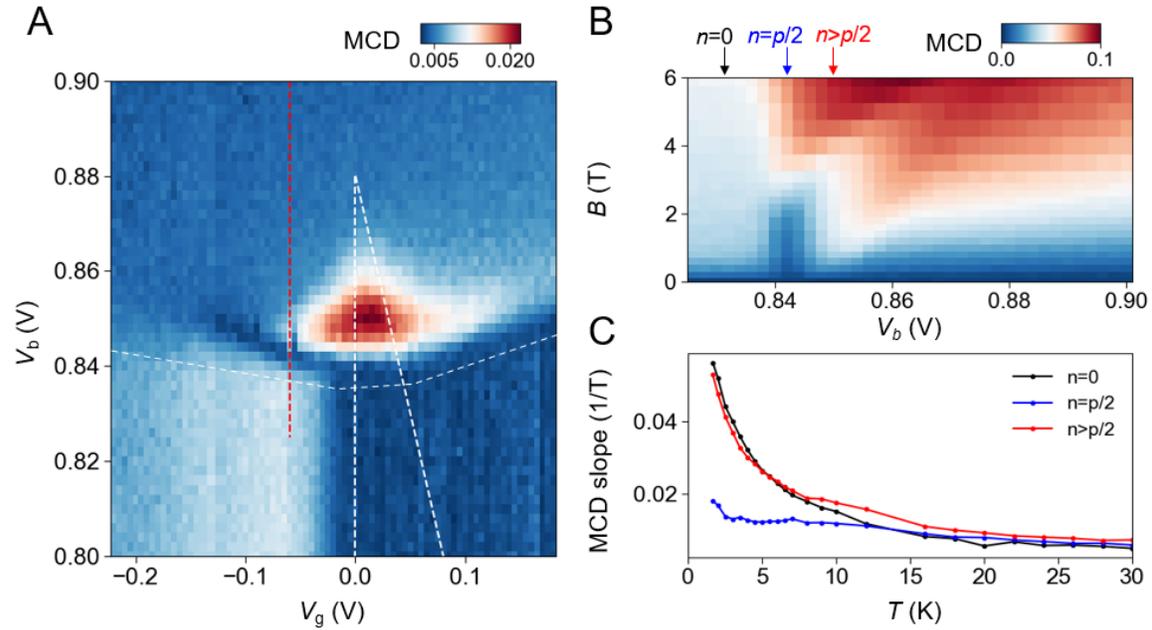

**Figure 3. Spin-singlet correlation in trions. A,** Spectrally integrated MCD of the W-layer as a function of $V_b$ and $V_g$ (device 3, out-of-plane magnetic field $B = 0.2$ T, temperature 1.6 K). White dashed lines denote the electrostatics phase boundaries. MCD is suppressed near the $p = 2n$ region. **B,** MCD versus $V_b$ and $B$ along the red dashed line in **A** (the corresponding net positive charge density is approximately $1.9 \times 10^{11}$ cm$^{-2}$). **C,** Temperature dependence of the zero-field MCD slope at $V_b = 0.83$ V, 0.842 V and 0.85 V, corresponding to $n = 0$, $= p/2$ and $> p/2$, respectively. The quantity is proportional to the magnetic susceptibility of the W-layer. Susceptibility at $p = 2n$ is suppressed below about 10 K.

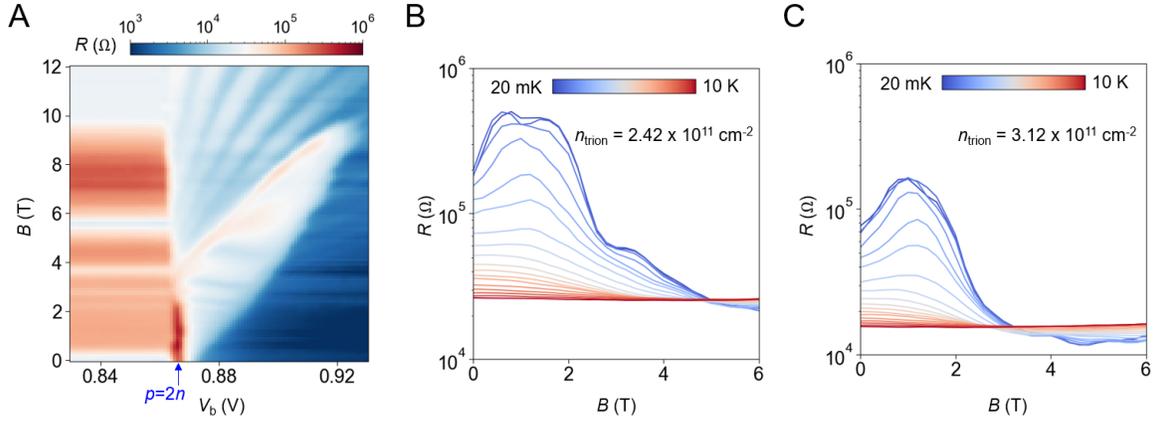

**Figure 4. Zeeman-field-induced dissociation of trions. A,** Four-terminal resistance $R$ versus $V_b$ and $B$ at constant $V_g$ (20 mK). The corresponding net positive charge density is approximately $2.4 \times 10^{11}$ cm$^{-2}$. **B,** Vertical linecut of **A** along $p = 2n$ at varying temperatures. The temperature dependence of $R$ is characteristic of an insulator below and of a metal above ~ 5 T. **C,** Same as **B** at a fixed trion density of $3.1 \times 10^{11}$ cm$^{-2}$. The critical field decreases with increasing trion density.

# Supplementary Materials for

## A degenerate trion liquid in atomic double layers

Phuong X. Nguyen, Raghav Chaturvedi, Liguo Ma, Patrick Knuppel, Kenji Watanabe, Takashi Taniguchi, Kin Fai Mak[*], Jie Shan[*]

[*]Corresponding authors. Email: jie.shan@cornell.edu; kinfai.mak@cornell.edu

**The PDF file includes:**

Materials and Methods

Supplementary Text

Figs. S1 to S8

References (*40-45*)

## Materials and Methods

### Device fabrication

The device geometry and contact design have been described in detail in Ref. (*17*). A schematic is illustrated in Fig. S1D, and the optical micrographs of the three devices are shown in Fig. S1A-C. The MoSe$_2$ and WSe$_2$ monolayers are embedded in the top and bottom graphite gates with hBN gate dielectrics of nearly identical thickness, $d_g$ = 10-20 nm. Platinum (Pt) and bismuth (Bi) are used as the metal electrodes for the W- and Mo-layers, respectively (*40, 41*). To achieve low contact resistance for charge injection and transport measurements, the devices are divided into three regions. In the channel region (I), the W- and Mo-layers are separated by a thin hBN spacer of thickness $d_{2L}$ = 1.5-2 nm (5-6 layers). In the contact region (II), the two TMDs are separated by a thicker hBN spacer (10-20 nm). Under a large displacement field ($\propto \Delta$), the TMDs in region III are heavily doped by one of the gates (specifically, a large negative top gate voltage for the WSe$_2$-Pt contact and a positive bottom gate voltage for the MoSe$_2$-Bi contact). Contact resistance down to about 10 kΩ below 1.5 K can be achieved. The intermediate region (II) is also heavily doped and serves as a charge reservoir for the channel.

The devices are fabricated using the layer-by-layer dry transfer technique as described in Ref. (*42*). The atomically thin flakes of van der Waals materials (TMDs, hBN and graphite) are mechanically exfoliated from their bulk crystals onto Si substrates with a 285-nm oxide layer. The flakes are sequentially picked up at 50°C using a polymer stamp made of a thin layer of polycarbonate on a polypropylene-carbonate-coated polydimethylsiloxane block. The completed stack after 9 transfers is released onto a Si substrate with pre-patterned Pt electrodes (5 nm Ti/25 nm Pt) to form contacts to the W-layer and to the graphite gate electrodes. The residual polycarbonate film is removed in chloroform and isopropanol. The contacts to the W-layer are completed by etching the top graphite gate using the standard electron-beam lithography and reactive-ion etching with oxygen plasma (Oxford Plasmalab80Plus). Last, the contacts to the Mo-layer are fabricated by a second-step electron beam lithography and metal deposition (thermal evaporation of Bi at a rate of 0.5 Å/s with a total thickness of 40-60 nm).

### Electrical measurements

The electrical measurements are performed using the standard lock-in technique in a Bluefors LD250 dilution refrigerator with lattice temperature down to about 20 mK. The contact electrode 1-5 to the W-layer and contact electrode 6-9 to the Mo-layer are labeled in Fig. S1A and B for device 1 and 2, respectively. The DC bias voltage $V_b$ is applied to electrode 6 of the Mo-layer while all other electrodes of the Mo-layer are floated. An AC bias voltage of 0.1 mV RMS (root mean square) at 7.33 Hz is applied to electrode 1, and the current is collected from electrode 5 of the W-layer. The bias current is kept below 5 nA at 20 mK to minimize sample heating. The voltage drop between electrode 2 and 3 is measured through a voltage preamplifier (Ithaco DL 1201) with input impedance 100 MΩ. The bias current and the voltage difference are simultaneously measured by lock-in amplifiers (Stanford Research SR830).

**Capacitance measurements**
The penetration capacitance and interlayer capacitance of the devices are measured by the lock-in technique to map the electrostatics phase diagram of the double layer (Fig. 1E). Details of the measurements have been described in Ref. (*15*). In short, for the penetration capacitance, an AC bias voltage of 5 mV (RMS) at 737 Hz is applied to the bottom gate. The induced carrier density change in the top gate is measured by a low-temperature capacitance bridge involving a GaAs high-electron-mobility-transistor (HEMT) amplifier (*43*). For the interlayer capacitance, an AC bias voltage of 5 mV (RMS) at 1.577 kHz is applied to the Mo-layer; the current is collected from the W-layer, amplified by a HEMT-based low-temperature transimpedance amplifier, and read by a lock-in amplifier (Stanford Research SR830). The in-phase component of the current is the tunneling current whereas the out-of-phase component (the displacement current) is proportional to the change in the pair density, $dn_{\text{pair}}$. The typical AC tunneling current is on the order of a few picoamps, in contrast to the in-plane bias current of ~ 5 nA (Fig. S8). The negligible tunneling currents support that the system is in the thermal equilibrium limit, and reliable in-plane transport measurements can be performed without artifacts from interlayer tunneling.

**Optical measurements**
The optical measurements are performed in a closed-cycle helium cryostat equipped with a 9 T superconducting magnet (Attocube, Attodry 2100, base temperature 1.6 K). A tungsten halogen lamp (Thorlabs) is used as a broadband white light source. Its output is coupled to a single mode fiber and focused onto the devices under normal incidence by a low temperature microscope objective (0.8 numerical aperture). The incident intensity on the samples is kept below 50 nW/$\mu m^{-2}$ to minimize the sample heating effect. The reflected light is spectrally resolved by a spectrometer coupled to a liquid nitrogen cooled silicon charge-coupled detector (Princeton Instruments). The reflectance contrast spectrum ($RC = \frac{R-R_0}{R_0}$) is obtained by comparing the reflected light spectrum, $R$, from the sample to a featureless background, $R_0$, obtained at high electron/hole doping densities. For polarization resolved measurements, the left ($\sigma^+$) and right ($\sigma^-$) circularly polarized light is generated by a combination of a linear polarizer and an achromatic quarter-wave plate. The MCD spectrum is defined as MCD = $(R^+ - R^-)/(R^+ + R^-)$, where $R^+$ and $R^-$ denote the reflected light spectrum of the left and right circularly polarized light, respectively.

Fig. S6A shows a representative reflectance contrast spectrum as a function of $V_b$ at fixed gate $V_g = 0.0$ V. For small bias voltages, the spectrum is dominated by the fundamental intralayer exciton resonance of the W-layer near 1.72 eV and the Mo-layer near 1.64 eV. At the threshold $V_b \approx 0.84$ V for exciton injection, the intralayer excitons lose their oscillator strengths abruptly, and the intralayer charged exciton features emerge near 1.70 eV nm and 1.61 eV, respectively. Fig. S6B is the MCD spectrum at $B = 0.2$ T. The absolute MCD value over the charged exciton feature for the W-layers (1.694 - 1.713 eV) and for the Mo-layer (1.600-1.621 eV) within the two vertical dashed lines is integrated and shown in Fig. 3 and Fig. S6C, respectively.

## Supplementary Text

**Determination of the $p = 2n$ line**

The effective circuit of the device (Fig. S2D) is three capacitors in series, where $C_g$ is the gate capacitance of the nearly symmetric top and bottom gates, and $C_I$ is the interlayer capacitance. The latter strongly depends on $V_b$, especially near the threshold of exciton injection. Only at large $V_b$, it approaches the geometrical value $C_{2L}$ of the electron-hole double layer. The differential change in the electron and hole densities ($dn$ and $dp$) can be expressed in terms of the differential change in voltages ($dV_b$ and $dV_g$):

$$dn \approx C_g(dV_g + dV_b) + C_I\, dV_b, \quad (1)$$
$$dp \approx -C_g dV_g + C_I dV_b. \quad (2)$$

The asymmetric part of the gate voltages, $\Delta$, is kept constant and thus its differential change $d\Delta = 0$. The $p = 2n$ line in the phase diagram satisfies $dp = 2dn$. Combining Eqns. 1 and 2, we obtain the local slope $\frac{dV_b}{dV_g}$ of the line in the electrostatics phase diagram

$$\frac{dV_b}{dV_g} \approx -3\left(2 + \frac{C_I}{C_g}\right)^{-1}, \quad (3)$$

which is fully determined by the capacitance ratio. The interlayer capacitance normalized by the geometric capacitance, $C_I/C_{2L}$, is measured as a function of $V_b$ and $V_g$ (Fig. S2B). The gate capacitance is calibrated from the slope of the Landau fans of the W-layer in the $V_b$- and $V_g$-map of the four-terminal resistance at large $V_b$ (red dashed line in the *pn* region, Fig. S2C). Along the line, the hole density is constant and hence $dp = 0$ or $\frac{dV_b}{dV_g} = \frac{C_g}{C_I}$ from Eqn. 1, which is approximately $\frac{C_g}{C_{2L}} \approx 0.15$. The value is independent of $V_b$ and $V_g$ in the limit of $C_g \ll C_I$ as in our experiment. The obtained value is consistent with the hBN dielectric layer thicknesses independently measured by the atomic force microscopy (AFM). The absolute value of $C_g$ can also be determined from the quantum oscillation period of the W-layer in the *pi* region of the electrostatics phase diagram (Fig. S2F).

The $p = 2n$ line (dashed line, Fig. S2B) starts at the lower left corner of the triangular phase of the excitonic insulator denoted by a blue dot (corresponding to $n = p = 0$). The local value of $C_I/C_{2L}$ is used to determine the slope $\frac{dV_b}{dV_g}$ (using Eqn. 3). We then propagate $V_b$ by a 2 mV step increase in $V_g$. We iterate the process by updating the local value of $C_I/C_{2L}$ after each step and obtain the $p = 2n$ line. The net charge particle density ($p - n$) along the line, which is also the trion density $n_{trion}$, can be determined by combining Eqns. 1 and 2, $p - n \approx -2C_g\Delta V_g - C_g\Delta V_b$, where $\Delta V_g$ and $\Delta V_b$ are the gate and bias voltages measured from the location of $n = p = 0$ (blue dot).

**Estimate of the trion Fermi temperature**

The trion Fermi temperature is given by $T_F = \frac{\pi \hbar^2 n_{trion}}{k_B m_{trion}}$, where $m_{trion}$, $\hbar$ and $k_B$ denote the trion effective mass, the Planck's constant and the Boltzmann constant, respectively. The trion effective mass is evaluated, $m_{trion} = 2m_h + m_e \approx 1.6\, m_0$, from the effective electron mass in the Mo-layer, $m_e \approx 0.8\, m_0$ [Ref. (44)], and the hole mass in the W-layer, $m_h \approx 0.4\, m_0$ [Ref. (45)] ($m_0$ denoting the bare electron mass). For a typical trion density, $n_{trion} = 2 \times 10^{11}$ cm$^{-2}$, the Fermi temperature is $T_F \approx 3.5$ K, which is substantially higher than the base temperature $T$ for the electrical measurements. The trion fluid is thus in the degenerate quantum limit $T \ll T_F$.

**Supplementary Figures**

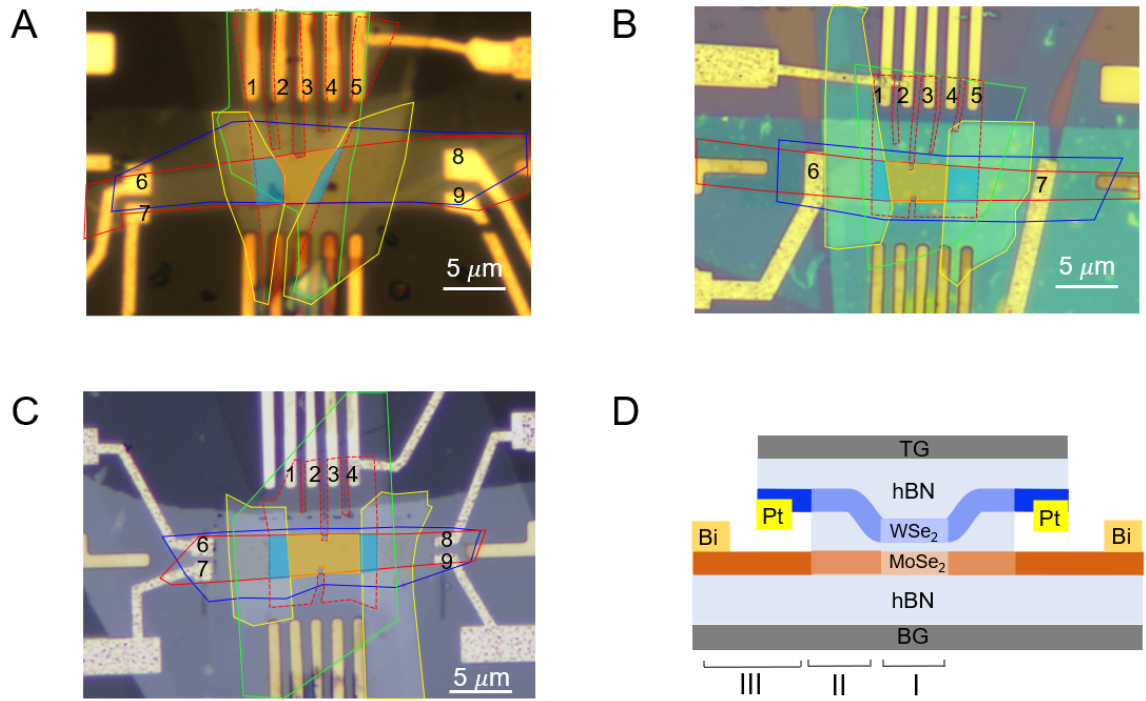

**Figure S1. Device images and schematic. A-C,** Optical micrographs for device 1, 2 and 3, respectively. The top gate (red dashed line), bottom gate (red line), W-layer (green line), Mo-layer (blue line) and the exciton contact hBN layers (yellow line) are outlined. The orange- and blue-shaded areas denote the channel and contact regions, respectively. The contact electrodes to the Mo- and W-layers are numbered. **D,** Cross section of the device showing the doping profile. Darker to lighter orange/blue color represents higher to lower electron/hole doping density. Region I is the channel, region II is the exciton contact and region III is the metal-TMD contact.

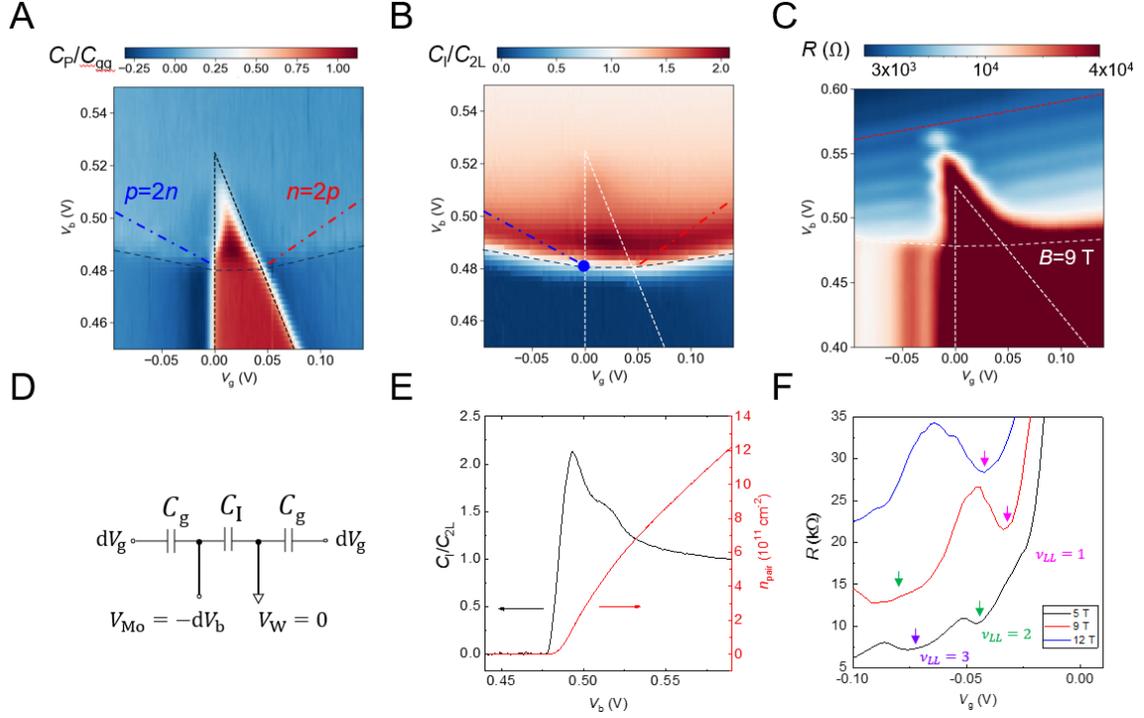

**Figure S2. Capacitance measurements and determination of the $p = 2n$ line. A, B,** Normalized penetration capacitance $C_P/C_{gg}$ (**A**) and interlayer capacitance $C_I/C_{2L}$ (**B**) as a function of $V_b$ and $V_g$ measured at $T = 1.5$ K (device 1). The dashed lines denote the W- and Mo-band edges and the exciton injection threshold; the blue and red dotted-dashed line denotes the $p = 2n$ and $n = 2p$ line, respectively (see Supplementary Text for its calibration). **C,** The four-terminal resistance of the W-layer as a function of $V_b$ and $V_g$ at $B = 9$ T and $T = 1.5$ K for device 1. The red dashed-line that traces along a specific Landau level in the W-layer at large $V_b$ is at a constant hole density; the slope of the line determines the ratio $C_g/C_{2L}$ (Supplementary Text). **D**. Effective capacitance circuit of the device with differential voltage $dV_g$ applied to the top and bottom gates, and $-dV_b$ applied to the Mo-layer while the W-layer is grounded. **E,** A line cut showing the dependence of $C_I/C_{2L}$ (black) and $n_{pair}$ (red) on $V_b$ at equal electron and hole densities. The pair density $n_{pair}$ is obtained by integrating $C_I$ with respect to $V_b$. **F,** The four-terminal resistance versus $V_g$ at varying magnetic fields taken at the *pi*-region in **C** ($V_b = 0.4$ V). The resistance minima of the quantum oscillations identify the Landau level indices. The distance between the adjacent minima, which corresponds to a single Landau level density, determines $C_g$ and the trion density along the $p = 2n$ line (Supplementary Text).

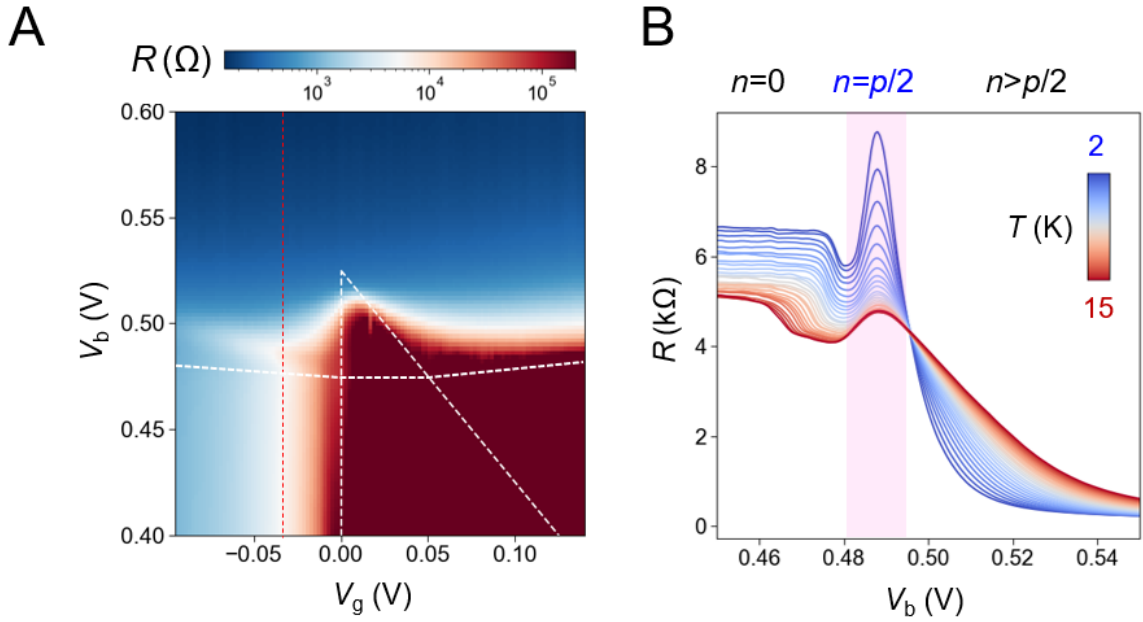

**Figure S3. Trion transport results for device 1. A,** Dependence of the four-terminal resistance $R$ ($T = 1.5$ K) on $V_g$ and $V_b$. **B,** Four-terminal resistance $R$ versus $V_b$ along the red dashed line in **A** at varying temperatures from 2 K to 15 K in 0.5 K step. An insulator-to-metal transition, which corresponds to a transition from a trion fluid at $n = p/2$ to an electron-hole plasma at $n > p/2$, is observed.

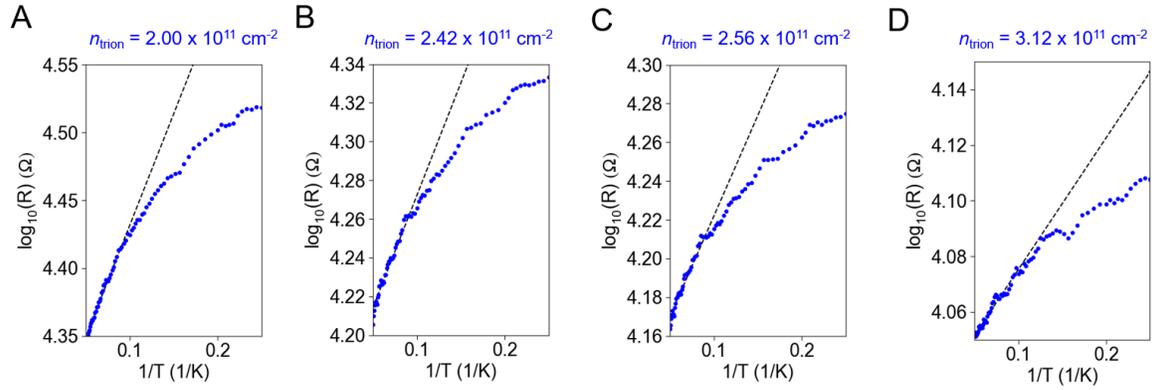

**Figure S4. Thermal activation fits to the temperature dependent resistance (device 2). A-D,** Arrhenius plot of the four-terminal resistance at varying trion densities along the $p = 2n$ line. The charge gaps are estimated by fitting the temperature dependence between 10 K and 20 K by a thermal activation dependence (dashed curves).

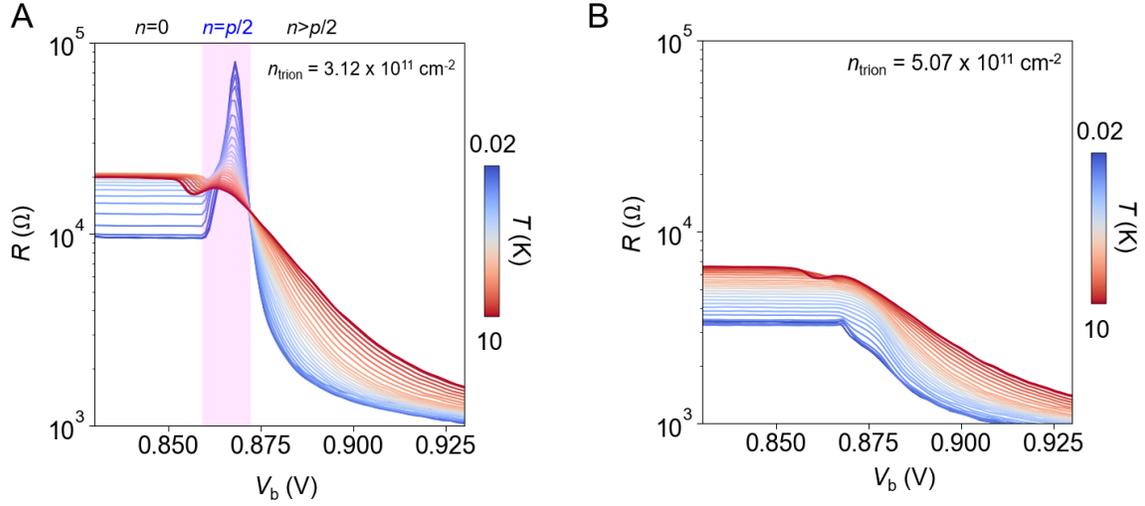

**Figure S5. Temperature dependent resistance for trion fluid and electron-hole plasma (device 2).** Four-terminal resistance versus $V_b$ at varying temperatures from 20 mK to 10 K at two additional $V_g$'s: one below (**A**, $V_g = -0.095$ V) and one above (**B**, $V_g = -0.165$ V) the critical value for the dissociation of trions. Whereas an insulating behavior is observed at $n = p/2$ in **A**, a metallic temperature dependence is observed irrespective of $V_b$ in **B**.

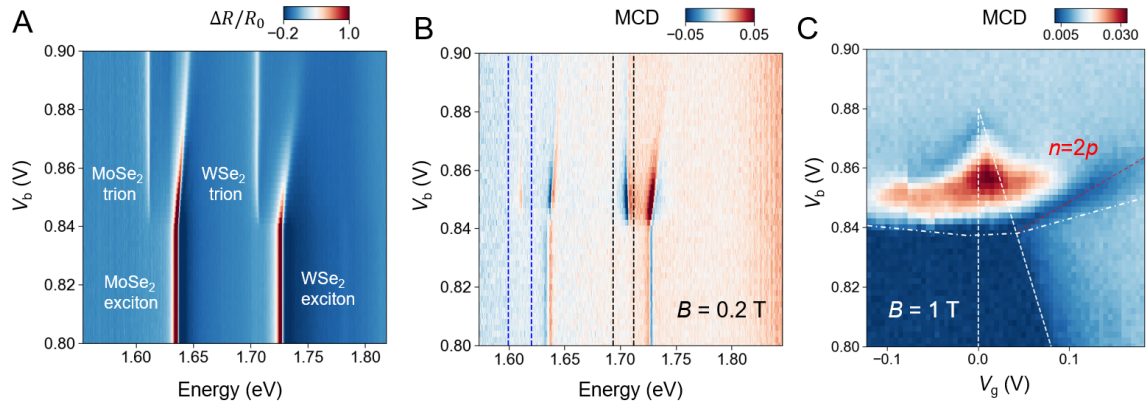

**Figure S6. Optical reflectance contrast and MCD spectra (device 3). A,** Reflection contrast spectrum as a function of $V_b$ at constant gate $V_g = 0.0$ V ($T = 1.6$ K and $B = 0$ T). Both the neutral and charged intralayer exciton resonances for both the TMD layers can be identified. **B,** MCD spectrum as a function of $V_b$ at constant gate $V_g = 0.0$ V ($T = 1.6$ K and $B = 0.2$ T). The dashed lines mark the spectral windows, which cover the charged exciton resonances of both TMD layers, we integrate to obtain the MCD signal for each TMD layer. **C,** Spectrally integrated MCD signal for the Mo-layer versus $V_g$ and $V_b$ at $B = 1$ T. Similar to the W-layer, there is a suppressed MCD (or suppressed magnetic susceptibility) along the $n \approx 2p$ line (red dashed line).

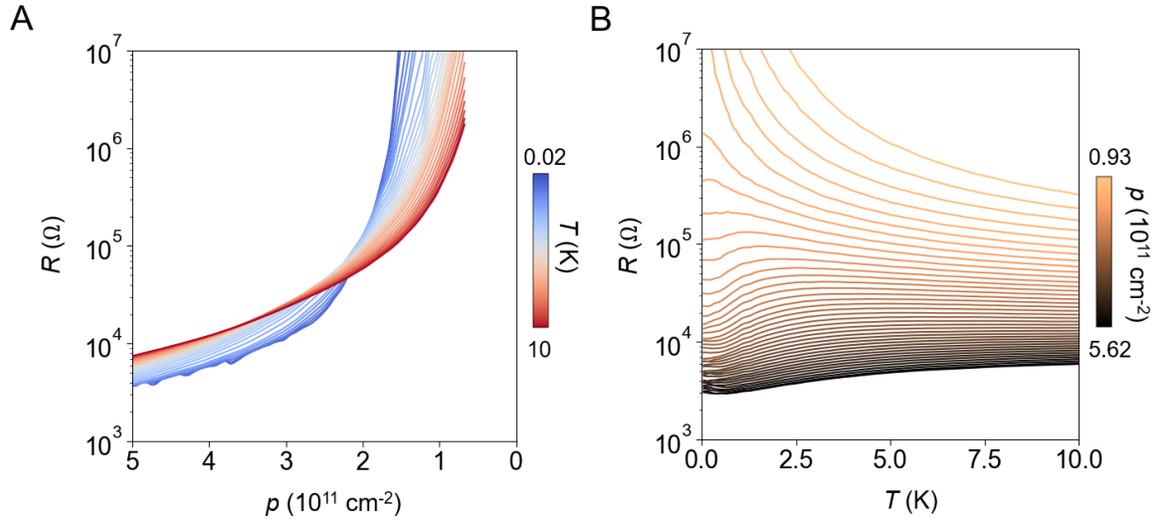

**Figure S7. Mobility edge for monolayer WSe$_2$ (device 2). A,** Four-terminal resistance as a function of the hole density in monolayer WSe$_2$ (in the *pi*-region) at varying temperatures from 20 mK to 10 K. A metal-to-insulator transition is observed at a critical density $p \approx 1.8 \times 10^{11}$ cm$^{-2}$. **B,** The corresponding temperature dependent resistance at varying hole densities from $0.93 \times 10^{11}$ to $5.62 \times 10^{11}$ cm$^{-2}$.

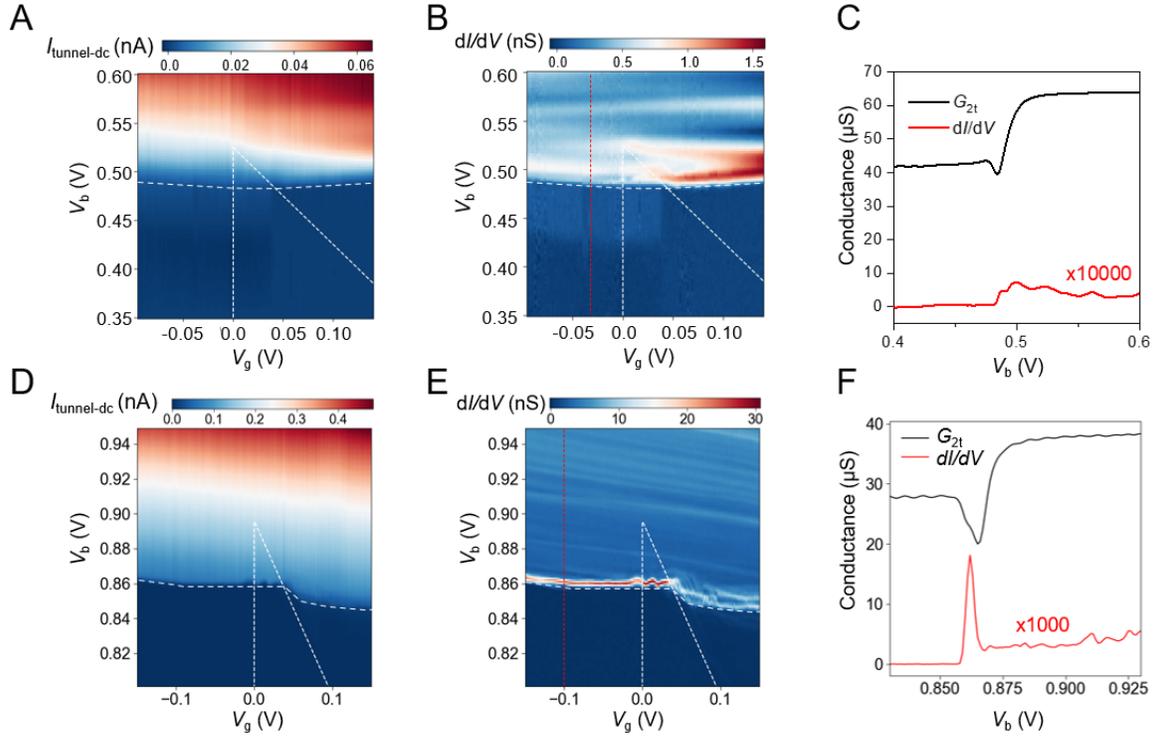

**Figure S8. Tunneling current for device 1 and 2.** DC tunneling current (**A**) and AC tunneling conductance (**B**) from W- to Mo-layer as a function of $V_b$ and $V_g$ for device 1. **C,** The AC tunneling conductance and the in-plane two-terminal conductance of the W-layer versus $V_b$ at constant $V_g$ (red dashed lines in **B**). The former is multiplied by a factor of 10,000. **D-F,** The same plots for device 2 are shown. Note that the bend on the electron side of the exciton injection threshold is caused by the large contact resistance in the Mo-layer in device 2. It does not affect the hole side we focus on in this study.